\documentstyle[11pt,epsbox]{article}

\newcommand{\mm}{\langle m_{\nu} \rangle_{\mu e}}

\title{{\bf Constraints of Mixing Angles from Lepton Number Violating Processes}}

\author{Hiroyuki NISHIURA \footnotemark[1],~~Kouichi MATSUDA \footnotemark[2]~~ 
and Takeshi FUKUYAMA  \footnotemark[2]  
                   \\
        $\ast$~~Junior College of Osaka Institute of Technology, \\
        Asahi-ku, Osaka, 535-8585 Japan \\
	$\dagger$~~Department of Physics, \\
        Ritsumeikan University, Kusatsu, \\
        Shiga, 525-8577 Japan}

\date{}

\begin{document}

\maketitle



\begin{abstract}
We discuss the constraints of lepton mixing angles from lepton number 
violating processes such as neutrinoless double beta decay, 
\(\mu^-\)-\(e^+\) conversion and K decay, $K^- \rightarrow \pi^+\mu^-\mu^-$  
which are allowed only if neutrinos are Majorana particles. The rates of 
these processes are proportional to the averaged neutrino mass
defined by $\langle m_{\nu} \rangle_{a b}\equiv |\sum _{j=1}^{3}U_{a j}
U_{b j}m_j|$ in the absence of right-handed weak coupling.
Here $a, b (j)$ are flavour(mass) eigen states and $U_{a j}$
is the left-handed lepton mixing matrix. We obtain the consistency 
conditions which are satisfied irrelevant to the concrete values of $CP$ 
violation phases (three phases in Majorana neutrinos). These conditions 
constrain the lepton mixing angles, neutrino masses $m_i$ and 
\(\langle m_{\nu} \rangle_{a b}\). By using these constraints we obtain 
the limits on the averaged neutrino masses for \(\mu^-\)-\(e^+\) conversion 
and K decay, $K^- \rightarrow \pi^+\mu^-\mu^-$.
\end{abstract}


\newpage



Recently it becomes more and more probable that neutrinos have masses.
This fact is due to the evidences of neutrino oscillation in a wide 
field such as solar neutrino oscillation \cite{kamioka} 
\cite{homestake} \cite{gallex} \cite{sage}, atmospheric neutrino deficit 
\cite{skamioka} and the neutrino oscillation from reactors 
\cite{chooz} \cite{bugey} and 
accelerators \cite{chorus} \cite{nomad} et al. 
Super-Kamiokande group \cite{skamioka} announced that they caught the 
definite evidence for $\nu_\mu$ oscillation in atmospheric neutrino deficit.
In these situations it is more and more indispensable to treat neutrino 
mixings from the wide variety of physics and to seek the consistency 
as a whole. At the present stage experimental information of 
$CP$ violating phases in leptonic sector is rather poor.  
Nevertheless, these phases may affect on the constraint of mixing 
angles and neutrino masses seriously. 
In the previous paper \cite{fuku} \cite{fuku2} it was shown that 
it is indeed the case and revealed how these phases constrain 
the mixing angles and masses irrespective to concrete values of 
$CP$ violating phases. By taking account of $CP$ violating phase effects 
in neutrinoless double beta decay seriously, we showed explicitly 
how it works in the confrontations with the neutrino oscillation data.

\par
In this paper we extend our arguments to the general lepton number 
violating processes which are allowed only if neutrinos are 
Majorana particles.
Besides neutrinoless double beta decay ($(\beta\beta)_{0\nu}$) 
(see Fig. 1a) discussed in the previous paper\cite{fuku}, \(\mu^-\)-\(e^+\) 
conversion process by the muonic atom, $\mu^-+N(A,Z+2) \rightarrow e^++N(A,Z)$  
\cite{takasugi} (see Fig.1b) and the lepton number violating K decay, 
$K^- \rightarrow \pi^+\mu^-\mu^-$  (see Fig.1c) are studied in the three 
generation case. In \cite{fuku} we discussed the relations between 
the mixing angles and 
neutrinoless double beta decay ($(\beta\beta)_{0\nu}$). 
The \(\mu^-\)-\(e^+\) conversion and K decay, 
$K^- \rightarrow \pi^+\mu^-\mu^-$ will give new information of 
the lepton mixing. The amplitudes of these three processes are, in the absence of 
right-handed weak couplings, proportional to the "averaged" mass 
$\langle m_{\nu} \rangle_{e e}$, $\langle m_{\nu} \rangle_{\mu e}$ 
and $\langle m_{\nu} \rangle_{\mu \mu}$ . The "averaged" mass 
$\langle m_{\nu} \rangle_{e e}$ defined from 
($(\beta\beta)_{0\nu}$) is
given by 
\begin{equation}
\langle m_{\nu} \rangle_{e e}\equiv |\sum _{j=1}^{3}U_{ej}^2m_j|, 
\label{eq761}
\end{equation}
The "averaged" mass $\langle m_{\nu} \rangle_{\mu e}$ defined from 
\(\mu^-\)-\(e^+\) conversion is,
\begin{equation}
\langle m_{\nu} \rangle_{\mu e}\equiv |\sum _{j=1}^{3}U_{\mu j}U_{e 
j}m_j|, \label{eq763}
\end{equation}
as pointed out in \cite{takasugi} and \cite{nishiura}. 
The "averaged" mass defined from the lepton number violating K decay,
$K^- \rightarrow \pi^+\mu^-\mu^-$ is 
\begin{equation}
\langle m_{\nu} \rangle_{\mu \mu}\equiv |\sum _{j=1}^{3}U_{\mu j}^2m_j|, \label{eq771}
\end{equation}
as pointed out in \cite{takasugi}. 
Here $U_{a j}$ is the left-handed lepton mixing matrix which combine 
the weak eigenstate neutrino ($a=e,\mu$ and $\tau$) to the mass 
eigenstate neutrino with mass $m_j$ ($j$=1,2 and 3).  In the case of Majorana 
neutrinos, $U$ takes the following form in the standard representation 
\cite{fuku}:
\begin{equation}
U=
\left(
\begin{array}{ccc}
c_1c_3&s_1c_3e^{i\beta}&s_3e^{i(\rho-\phi )}\\
(-s_1c_2-c_1s_2s_3e^{i\phi})e^{-i\beta}&
c_1c_2-s_1s_2s_3e^{i\phi}&s_2c_3e^{i(\rho-\beta )}\\
(s_1s_2-c_1c_2s_3e^{i\phi})e^{-i\rho}&
(-c_1s_2-s_1c_2s_3e^{i\phi})e^{-i(\rho-\beta )}&c_2c_3\\
\end{array}
\right).\label{eq772}
\end{equation}
Here $c_j=\cos\theta_j$, $s_j=\sin\theta_j$ 
($\theta_1=\theta_{12},~\theta_2=\theta_{23},~\theta_3=\theta_{31}$). 
Three 
$CP$ violating phases, $\beta$ , $\rho$ and $\phi$ appear in $U$ for 
Majorana
particles \cite{bilenky}.
\par
Experimental data  for $\langle m_{\nu} \rangle_{e e}$, the branching ratio 
of the \(\mu^-\)-\(e^+\) conversion to $\mu$ capture, and the branching ratio of 
K decay, $K^- \rightarrow \pi^+\mu^-\mu^-$ are as follows. \cite{pd}

\begin{eqnarray}
\langle m_{\nu} \rangle_{e e} < (0.5-1.5) eV,
\end{eqnarray}
\begin{eqnarray}
{ \sigma (\mu ^-{}^{32}S\rightarrow e^+{}^{32}Si^*) \over \sigma (\mu 
^-{}^{32}S\rightarrow \nu_{\mu}{}^{32}P^*)}<9\times 10^{-10}, \label{eq773}
\end{eqnarray}
\begin{eqnarray}
{\Gamma(K^- \rightarrow \pi^+\mu^-\mu^-) \over \Gamma(K^- \rightarrow all)}
 <1.5\times 10^{-4}.\label{eq774}
\end{eqnarray}

The experimental data lead to rather large upper bounds for $\langle m_{\nu} 
\rangle_{\mu e}$ and $\langle m_{\nu} \rangle_{\mu \mu}$ at present 
except for $\langle m_{\nu} \rangle_{e e}$. 
Using theoretical estimations\cite{takasugi}, 
one obtain from Eq.\ (\ref{eq773}) that  
$\langle m_{\nu} \rangle_{\mu e} < 2$ GeV or $400$ GeV for the spin 
factor of daughter nuclei \(S=1\) or \(S=-1\), respectively. One also obtain from 
Eq.\ (\ref{eq774}) that $\langle m_{\nu} \rangle_{\mu \mu} < 1\times 10^{5}$ GeV.  
In this paper we derive the consistency conditions among the lepton mixing angles, 
neutrino mass $m_i$ and averaged neutrino mass $\langle m_{\nu} 
\rangle_{a b}$. From those constraints we try to get the information of 
\(\langle m_{\nu} \rangle_{\mu e}\) and \(\langle m_{\nu} \rangle_{\mu \mu}\).

\par
We now discuss the relations between lepton mixing angles and "averaged" mass 
$\langle m_{\nu} \rangle_{a b}$ ($a,b$ = $\mu$ or e). 
Substitution of the expression Eq.(\ref{eq772}) 
into Eqs.(\ref{eq761})-(\ref{eq771}) shows that $\langle 
m_{\nu} \rangle_{\mu e}$ and $\langle m_{\nu} \rangle_{\mu \mu}$ are 
functions of three $CP$ violating phases, 
whereas $m_{\nu} \rangle_{e e}$ in ($\beta\beta)_{0\nu}$ was a function 
of the two $CP$ violating phases.  However, we can treat these three cases, 
and the other cases if needed, equally well by assuming that the $CP$ violating 
phase $\phi$ is known. 
Indeed, $\phi$ is detectable 
in neutrino oscillation processes, whereas $\beta$ and $\rho$ are not.
Therefore, we first derive, from the expression of the "averaged" mass, 
the consistency conditions which are irrelevant to the concrete values of 
two $CP$ violating phases $\beta$ and $\rho$.
These consistency conditions are relations among the neutrino masses, 
mixing angles, $\langle m_{\nu} \rangle_{a b}$ and $CP$ violating phase $\phi$. 
First, we explain this and proceed to study $\langle m_{\nu} \rangle_{a b}$.

\par
The "averaged" mass $\langle m_{\nu} \rangle_{a b}$ defined by 
Eqs.(\ref{eq761})-(\ref{eq771}) can be expressed as
\begin{equation}
\langle m_{\nu} \rangle_{a b}\equiv |\sum _{j=1}^{3}U_{a j}
U_{b j}m_j|=|X_1+X_2e^{2i\xi}+X_3e^{2i\eta}|,
\end{equation}
where $X_i$ ($i$=1$\sim$3) are positive numbers which depend on mixing angles 
and $CP$ violating phase $\phi$. $\xi$ and $\eta$ are $CP$ violating phases 
dependent on $\beta$ and $\rho$, which are characteristic of 
Majorana neutrinos.  So we obtain
\begin{equation}
\langle m_{\nu} \rangle_{a b}^2=X_1^2+X_2^2+X_3^2+2X_1X_2\cos 
2\xi +2X_1X_3\cos 2\eta +2X_2X_3\cos (2\xi-2\eta ). \label{eq775}
\end{equation}
Using $\cos 2\xi={1-\tan^2\xi\over 1+\tan^2\xi}$ and $\sin 
2\xi={2\tan \xi\over 1+tan^2\xi}$, Eq.(\ref{eq775}) is rewritten as
\begin{eqnarray}
\{(X_1-X_2)^2+2(X_1-X_2)X_3\cos 2\eta +X_3^2-\langle m_{\nu} 
\rangle_{a b}^2\}\tan^2\xi+4X_2X_3\sin 2\eta\tan\xi \nonumber \\
+\{(X_1+X_2)^2+2(X_1+X_2)X_3\cos 2\eta +X_3^2-\langle m_{\nu} 
\rangle_{a b}^2\}=0. \label{eq7706}
\end{eqnarray}
We regard this equation as a quadratic equation of $\tan\xi$ and its 
positive discriminant indicates that
\begin{equation}
{(\langle m_{\nu} \rangle_{a b}-X_2)^2-(X_1-X_3)^2\over 4X_1X_3}
\le \cos^2\eta\le {(\langle m_{\nu} 
\rangle_{a b}+X_2)^2-(X_1-X_3)^2\over 4X_1X_3}. \label{eq7707}
\end{equation}
In the above process Eqs.(\ref{eq7706})-(\ref{eq7707}) we may 
replace the role of $\xi$ by $\eta$. 
Namely expressing $\cos 2\eta$ and $\sin 2\eta$ in terms of $\tan\eta$, 
we obtain 
\begin{equation}
{(\langle m_{\nu} \rangle_{a b}-X_3)^2-(X_1-X_2)^2\over 4X_1X_2}
\le \cos^2\xi\le {(\langle m_{\nu} 
\rangle_{a b}+X_3)^2-(X_1-X_2)^2\over 4X_1X_2} \label{eq7708}
\end{equation}
Neutrino oscillation processes do not depend on $\beta$ and $\rho$. 
So these two inequalities become the constraint on the $CP$ violating phases 
if the mixing angles and eigen masses will be pursued from oscillation 
experiments. However, at present we want to know the constraints on the 
angles and masses which are irrelevant to the specific values of the $CP$ 
violating phases. Those are followed from Eqs. (\ref{eq7707}) and 
(\ref{eq7708}) by remarking 
$0\le \cos^2\eta\le 1$ and $0\le \cos^2\xi\le 1$ . Namely we find 
\begin{equation}
B\le \langle m_{\nu} \rangle_{a b},\qquad C\le \langle m_{\nu} \rangle_{a b},
\qquad D\le \langle m_{\nu} \rangle_{a b},\qquad 
\langle m_{\nu} \rangle_{a b}\le A.
\end{equation}
which we denote as 
\begin{equation}
\mbox{Max} \{B,C,D\} \le \langle m_{\nu} \rangle_{a b}\le A.
\label{eq7709}
\end{equation}
Here
\begin{equation}
A\equiv X_1+X_2+X_3,\qquad B\equiv X_1-X_2-X_3,\qquad C\equiv 
X_2-X_3-X_1,\qquad D\equiv X_3-X_1-X_2.
\label{eq7710}
\end{equation}
It should be noticed that inequality (\ref{eq7709}) is symmetric with respect to 
$X_i$.
 Eqs.(\ref{eq7707})-(\ref{eq7709}) 
are the general expression of the constraint appeared in 
$(\beta\beta)_{0\nu}$, \(\mu^-\)-\(e^+\) conversion and 
$K^- \rightarrow \pi^+\mu^-\mu^-$.  Hereafter we 
proceed to discuss the specific process.
\par
In the case of $(\beta\beta)_{0\nu}$, 
it follows from Eqs. (\ref{eq771}) and (\ref{eq772}) 
that $X_i$, $\xi$ and $\eta$ become
\begin{equation}
X_1=m_1c_1^2c_3^2,\quad X_2=m_2s_1^2c_3^2, \quad X_3=m_3s_3^2
\end{equation}
and
\begin{equation}
\xi=\beta, \qquad \eta=\rho -\phi\equiv \rho '.
\end{equation}
Then Eq. (\ref{eq7709}) coincides with the results (2.11) and (2.14) in 
\cite{fuku}.  
Without loss of generality we may set $m_1\le m_2\le m_3$. 
Then Eq.(\ref{eq7709}) 
reveals useful information explicitly.  For instance, using $m_1\le 
m_2$, we obtain
\begin{equation}
m_3s_3^2-m_2(1-s_3^2)\le D\le \langle m_{\nu} \rangle_{ee}\le A\le 
m_3s_3^2+m_2(1-s_3^2).
\end{equation}
which leads to upper bound of $s_3^2$ as
\begin{equation}
s_3^2 \le {m_2+\langle m_{\nu} \rangle_{e e}\over m_3+m_2} 
\end{equation}
This was an important constraint in $(\beta\beta)_{0\nu}$ 
\cite{fuku}.  Eq.(\ref{eq7709}) offers other information.

\par
In the case of \(\mu^-\)-\(e^+\) conversion, the averaged mass is 
\begin{eqnarray}
\langle m_{\nu} \rangle_{\mu e}& = & 
|m_1c_1c_3R_1(\phi)e^{-i(\beta-\chi_1)}+
m_2s_1c_3R_2(\phi)e^{i(\beta+\chi_2)}+m_3s_2
s_3c_3e^{i(2\rho-\beta-\phi)}| \nonumber \\
& = & |m_1c_1c_3R_1(\phi)+m_2s_1c_3R_2(\phi)e^{2i\xi}+m_3s_2s_3c_3e^{2i\eta}|
 \label{mat291}\\
& \equiv & |X_1+X_2e^{2i\xi}+X_3e^{2i\eta}|. \nonumber
\end{eqnarray}
Here $R_a$ and $\chi_a$ $(a=1,2)$ are defined in terms of $U_{\mu i}$ in 
Eq.(\ref{eq772})
\begin{equation}
U_{\mu 1}\equiv R_1(\phi)e^{i\chi_1}e^{-i\beta},
 \quad U_{\mu 2}\equiv R_2(\phi)e^{i\chi_2}
\end{equation}
and
\begin{equation}
2\xi = 2\beta-\chi_1+\chi_2, \qquad 2\eta = 2\rho-\phi-\chi_1.
\end{equation}
The explicit forms of $R_a$ and $X_i$ are therefore
\begin{eqnarray}
R_1(\phi)^2 & = & s_1^2c_2^2+2s_1s_2s_3c_1c_2\cos\phi 
+c_1^2s_2^2s_3^2,\nonumber \\
R_2(\phi)^2 & = & c_1^2c_2^2-2s_1s_2s_3c_1c_2\cos\phi +s_1^2s_2^2s_3^2,
\label{eq7717}\\
S \hspace{0.5cm} 
 &\equiv& R_1(\phi)^2 + R_2(\phi)^2 = 1-s_2^2c_3^2, \nonumber
\end{eqnarray}
and
\begin{equation}
X_1=m_1c_1c_3R_1(\phi)\equiv x_1R_1(\phi),\quad X_2=m_2s_1c_3R_2(\phi)
\equiv x_2R_2(\phi), \quad 
X_3=m_3s_2s_3c_3.\label{eq7718}
\end{equation}
That is, $X_1$ and $X_2$ are, contrary to the case of 
$(\beta\beta)_{0\nu}$, dependent on the phase $\phi$. This phase appears 
and is detectable in neutrino oscillation processes in general.  So Eq. (\ref{eq7709}) 
gives the constraint among the neutrino masses, mixing angles, 
$\langle m_{\nu} \rangle_{\mu e}$ and the phase $\phi$.

\par
In the case of K decay, $K^- \rightarrow \pi^+\mu^-\mu^-$ , 
the averaged mass is 
\begin{eqnarray}
\langle m_{\nu} \rangle_{\mu \mu}& = & 
|m_1R_1^2(\phi)e^{-2i(\beta-\chi_1)}+
m_2R_2^2(\phi)e^{2i\chi_2}+m_3s_2^2
c_3^2e^{2i(\rho-\beta)}| \nonumber \\
& = & |m_1R_1^2(\phi)+m_2R_2^2(\phi)e^{2i\xi}+m_3s_2^2c_3^2e^{2i\eta}|
 \\
& \equiv & |X_1+X_2e^{2i\xi}+X_3e^{2i\eta}|. \nonumber
\end{eqnarray}
Here 
\begin{equation}
\xi = \beta-\chi_1+\chi_2, \qquad \eta = \rho-\chi_1.
\end{equation}
The explicit forms of $X_i$ are therefore
\begin{equation}
X_1=m_1R_1^2(\phi),\quad X_2=m_2R_2^2(\phi), \quad X_3=m_3s_2^2c_3^2.\label{eq7719}
\end{equation}
That is, $X_1$ and $X_2$ also depend on the phase $\phi$. So Eq. (\ref{eq7709}) 
gives the constraint among the neutrino masses, mixing angles, 
$\langle m_{\nu} \rangle_{\mu \mu}$ and the phase $\phi$.

\par
Now we discuss the consistency conditions Eq.(\ref{eq7709}) more explicitly for 
neutrinoless double beta decay, \(\mu^-\)-\(e^+\) conversion and K decay, 
$K^- \rightarrow \pi^+\mu^-\mu^-$. 
Recent neutrino oscillation experiments suggest that $\theta_{13}$ is 
very small, $s_3^2 < 0.05$ \cite{chooz} and we set it approximately to be zero. 
In this case the phase $\phi$ disappears and 
the consistency condition of $(\beta\beta)_{0\nu}$ becomes
\begin{equation}
{\langle m_{\nu} \rangle_{e e}-m_1 \over m_2-m_1}\leq 
s_1^2,\quad {m_1-\langle m_{\nu} \rangle_{e e} \over m_2+m_1}\leq 
s_1^2,\quad s_1^2\leq { \langle m_{\nu} \rangle_{e e}+m_1 \over 
m_1+m_2}. 
\end{equation}
which leads to the allowed bound on $s_1$.
The consistency condition of \(\mu^-\)-\(e^+\) conversion becomes
\begin{equation}
{\langle m_{\nu} \rangle ^2_{\mu e}\over (m_1+m_2)^2}\leq 
s_1^2c_1^2c_2^2\leq {\langle m_{\nu} \rangle ^2_{\mu e}\over 
(m_1-m_2)^2}.\label{eq7802}  
\end{equation}
The inequality gives us the allowed region on 
$\sin^22\theta_1$ - $\langle m_{\nu} \rangle_{\mu e} ^2$ plane ,which is depicted 
by the shaded areas in Fig.2. 
From Fig.2 it is found that $\langle m_{\nu} \rangle_{\mu e} ^2$ is bounded as 
${(m_2-m_1)^2c_2^2\over 4}\leq \langle m_{\nu} \rangle ^2_{\mu e}
\leq {(m_2+m_1)^2c_2^2\over 4}$. 
The inequality Eq.\ (\ref{eq7802}) also gives us the allowed region on 
$s_1^2$ - $s_2^2$ plane ,which is depicted by the shaded areas in Fig.3. 
In order for allowed region to exist we find, from Fig.3, that
\begin{equation}
{\langle m_{\nu} \rangle_{\mu e} \le {m_1+m_2 \over 2}}.
\end{equation}
So we predict the averaged mass 
$\langle m_{\nu} \rangle_{\mu e}$ is at most ${(m_2+m_1)\over 2}$.
From Fig.3, we also obtain the following bounds on $s_1^2$ :
\begin{equation}
{1\over 2}-{\sqrt{(m_2-m_1)^2-4\langle m_{\nu} \rangle ^2_{\mu e}}\over 2(m_2-m_1)}
\leq s_1^2 \leq {1\over 2}+{\sqrt{(m_2-m_1)^2-4\langle m_{\nu} \rangle ^2_{\mu e}}
\over 2(m_2-m_1)}. 
\end{equation}
The consistency condition of K decay, $K^- \rightarrow \pi^+\mu^-\mu^-$ 
becomes
\begin{eqnarray}
{-m_2-\langle m_{\nu} \rangle_{\mu \mu}+(m_2+m_1)s_1^2 \over m_3-m_2+(m_2+m_1)s_1^2}\leq 
s_2^2, \nonumber \\
{m_2-\langle m_{\nu} \rangle_{\mu \mu}-(m_2+m_1)s_1^2 \over m_3+m_2-(m_2+m_1)s_1^2}\leq 
s_2^2, \nonumber \\
{-m_2+\langle m_{\nu} \rangle_{\mu \mu}+(m_2-m_1)s_1^2 \over m_3-m_2+(m_2-m_1)s_1^2}\leq 
s_2^2, \\
s_2^2\leq { m_2+\langle m_{\nu} \rangle_{\mu \mu}-(m_2-m_1)s_1^2 \over 
m_3+m_2-(m_2-m_1)s_1^2}.\nonumber  
\end{eqnarray}
This inequality gives us the allowed region on $s_1^2$ - $s_2^2$ plane. 
The allowed region is given in Fig.4 by the shaded areas in the cases: 
\ (a) $\langle m_\nu \rangle_{\mu \mu} \leq m_1$, 
\ (b) $m_1 \leq \langle m_\nu\rangle_{\mu \mu} \leq m_2 $ and
\ (c) $m_2 \leq \langle m_\nu \rangle_{\mu \mu} \leq m_3$. 
From Fig.4, we obtain the following bounds on $s_2^2$ :
\begin{eqnarray}
s_2^2\leq \frac{m_2+\langle m_{\nu} \rangle_{\mu \mu}}{m_3+m_2} 
\hspace{2cm} &\mbox{for}& \qquad\langle m_\nu \rangle_{\mu \mu}
 \leq m_2 \nonumber \\\frac{\langle m_{\nu}
 \rangle_{\mu \mu}-m_2}{m_3-m_2} \le s_2^2 
\leq \frac{m_2+\langle m_{\nu} \rangle_{\mu \mu}}{m_3+m_2} 
\qquad &\mbox{for}& \qquad m_2 \le \langle m_\nu \rangle_{\mu \mu} \leq m_3. 
\end{eqnarray}  
This inequality gives us the allowed region $s_3^2$ versus 
$\langle m_{\nu} \rangle_{\mu \mu}$ plane ,which is shown by 
the shaded areas in Fig.5. From Fig.5 we find 
\begin{equation}
(m_3+m_2)s_2^2-m_2 \le \langle m_{\nu} \rangle_{\mu \mu} \le (m_3-m_2)s_2^2+m_2.
\end {equation} 
So we predict the limit for $\langle m_{\nu} \rangle_{\mu \mu}$ as ${m_3-m_2 \over2}
\le \langle m_{\nu} \rangle_{\mu \mu} \le {m_3+m_2 \over2}$  for maximal mixing of $\theta_2$.

\par
We next discuss the elimination of $CP$ violation phase from the 
constraint Eq.(\ref{eq7709}). 
In the general case where $s_3$ is not zero, Eq.(\ref{eq7709}) includes 
the phase $\phi$ for the cases of \(\mu^-\)-\(e^+\) conversion and 
$K^- \rightarrow \pi^+\mu^-\mu^-$. If we want to 
know the constraints only among the neutrino masses, mixing angles and  
$\langle m_{\nu} \rangle_{a b}$, we must eliminate $\phi$. This is performed as follows.
At first let us discuss the constraints for \(\mu^-\)-\(e^+\) conversion. 
By remarking $-1 \le \cos\phi \le 1$, we find, 
from Eqs. (\ref{eq7709}), (\ref{eq7710}) and (\ref{eq7718}),
the constraint condition only among the neutrino masses, mixing angles and  
$\langle m_{\nu} \rangle_{\mu e}$ as
\begin{equation}
 \mbox{Max}
 \left\{
  \begin{array}{c}
    -\sqrt{S(x_1^2+x_2^2)}+X_3\\
    x_1 R_{1-}-x_2 R_{2+}-X_3\\
    -x_1 R_{1+}+x_2 R_{2-}-X_3\\
  \end{array}\right\}
 \le \mm \le
  \sqrt{S(x_1^2+x_2^2)}+X_3, \label{eq8801}  
\end{equation}
where  \(X_3\), \(x_1\) and 
\(x_2\) are defined by Eq.(\ref{eq7718}) and \(R_{i \pm}\)($i=1,2$) are defined by
\begin{equation}
R_{1 \pm} \equiv  |s_1c_2 \pm c_1s_2s_3|, \qquad R_{2 \pm} \equiv  |c_1c_2 \pm s_1s_2s_3|.
\end{equation}
It also should be noted that from the the constraints Eqs. (\ref{eq7709}), (\ref{eq7710}) and (\ref{eq7718}), 
we have bounds for $cos\phi$ as 
\begin{eqnarray}
\frac{f_1^2 - s_1^2 c_2^2 - s_2^2 s_3^2 c_1^2}{2s_1 s_2 s_3 c_1 c_2}
&\le \cos \phi \le&
\frac{g_1^2 - s_1^2 c_2^2 - s_2^2 s_3^2 c_1^2}{2s_1 s_2 s_3 c_1 c_2}, \label{241}\\
\frac{c_1^2 c_2^2 + s_1^2 s_2^2 s_3^2 - g_2^2}{2s_1 s_2 s_3 c_1 c_2}
&\le \cos \phi \le&
\frac{c_1^2 c_2^2 + s_1^2 s_2^2 s_3^2 - f_2^2}{2s_1 s_2 s_3 c_1 c_2}.\label{242}
\end{eqnarray}
where \(f_1\) and \(g_1\) are defined by
\begin{eqnarray}
f_1 &\equiv& \mbox{Max}
\left\{
\begin{array}{c}
 \frac{x_1(\mm-X_3)-x_2\sqrt{S(x_1^2+x_2^2)-(\mm-X_3)^2}}{x_1^2+x_2^2},\\
 \frac{x_1(X_3-\mm)-x_2\sqrt{S(x_1^2+x_2^2)-(\mm-X_3)^2}}{x_1^2+x_2^2},\\
 \frac{-x_1(\mm+X_3)+x_2\sqrt{S(x_1^2+x_2^2)-(\mm+X_3)^2}}{x_1^2+x_2^2},\\
 0
\end{array}
\right\},\\
g_1 &\equiv& \mbox{Max}
\left\{
\begin{array}{c}
 \frac{x_1(\mm+X_3)+x_2\sqrt{S(x_1^2+x_2^2)-(\mm+X_3)^2}}{x_1^2+x_2^2},\\
 \frac{X_3+\mm}{x_1}
\end{array}
\right\}.
\end{eqnarray}
The \(f_2\)(\(g_2\)) is defined by 
the exchange of \(x_1\) for \(x_2\) in \(f_1\)(\(g_1\)). 
\par
Next let us discuss the constraints Eqs. (\ref{eq7709}), (\ref{eq7710}) and (\ref{eq7719}) 
of $K^- \rightarrow \pi^+\mu^-\mu^-$.
By the similar method in  \(\mu^-\)-\(e^+\) conversion, we find the following constraint 
which is irrelevant to concrete value of $\phi$ and the bound of $cos\phi$:
\begin{equation}
\mbox{Max}
\left\{
 \begin{array}{c}
  -m_1 R_{1+}^2 + m_2 R_{2-}^2-m_3s_2^2c_3^2\\
   m_1 R_{1-}^2 - m_2 R_{2+}^2-m_3s_2^2c_3^2\\
  -m_1 R_{1-}^2 - m_2 R_{2+}^2+m_3s_2^2c_3^2
 \end{array}\right\}
\le \langle m_\nu \rangle_{\mu\mu} \le
m_1 R_{1-}^2 + m_2 R_{2+}^2+m_3s_2^2c_3^2.
\end{equation}
and 
\begin{eqnarray}
&&\frac{-m_1(s_1^2c_2^2+c_1^2s_2^2s_3^2)+m_2(c_1^2c_2^2+s_1^2s_2^2s_3^2)-m_3s_2^2c_3^2-\langle
m_\nu \rangle_{\mu\mu}}
     {2(m_1+m_2)c_1c_2s_1s_2s_3} \le \cos \phi \le  \\
&& \hspace{3cm}
\mbox{Min}
\left\{
 \begin{array}{c}
  \frac{m_1(s_1^2c_2^2+c_1^2s_2^2s_3^2)+m_2(c_1^2c_2^2+s_1^2s_2^2s_3^2)+m_3s_2^2c_3^2-\langle m_\nu \rangle_{\mu\mu}}
       {2(m_2-m_1)c_1c_2s_1s_2s_3}\\
  \frac{-m_1(s_1^2c_2^2+c_1^2s_2^2s_3^2)+m_2(c_1^2c_2^2+s_1^2s_2^2s_3^2)+m_3s_2^2c_3^2+\langle m_\nu \rangle_{\mu\mu}}
       {2(m_1+m_2)c_1c_2s_1s_2s_3}\\
  \frac{m_1(s_1^2c_2^2+c_1^2s_2^2s_3^2)+m_2(c_1^2c_2^2+s_1^2s_2^2s_3^2)-m_3s_2^2c_3^2+\langle m_\nu \rangle_{\mu\mu}}
       {2(m_2-m_1)c_1c_2s_1s_2s_3}\nonumber
 \end{array}\right\}
\end{eqnarray}

\par
We finally discuss the constraints for $CP$ violating phase $\beta$ and 
the constraint which is irrelevant to concrete value of $\beta$ 
from the \(\mu^-\)-\(e^+\) conversion .
Among the three phases, Dirac phase $\phi$ is of primary importance since 
it survives in neutrino oscillation processes in general. 
However if want to know the 
constraint for Majorana phase $\beta$, we may transform Eq. 
(\ref{eq763}) as
\begin{eqnarray}
\mm&=&|(c_2c_3)( -m_1s_1c_1 + m_2s_1c_1 e^{2i\beta})
+(s_2s_3c_3)(-m_1c_1^2-m_2s_1^2e^{2i\beta})e^{i\phi}\nonumber\\
&& +m_3s_2s_3c_3e^{i(2\rho-\phi)}|\nonumber\\
&=& |c_2c_3R_1'(\beta)+s_2s_3c_3R_2'(\beta) e^{2i \xi'}+ 
    m_3s_2s_3c_3e^{2i\eta'}|\\
&=& |x_1'R_1'(\beta)+x_2'R_2'(\beta)e^{2i \xi'}+X_3'e^{2i\eta'}|. \nonumber
\end{eqnarray}
where  $R_a'(\beta)$ and $\chi_a'$ $(a=1,2)$ are defined by 
\begin{equation}
-m_1s_1c_1+m_2s_1c_1e^{2i\beta}\equiv R_1'(\beta)e^{i\chi_1'},
 \quad -m_1c_1^2-m_2s_1^2e^{2i\beta}\equiv R_2'(\beta)e^{i\chi_2'}
\end{equation}
and
\begin{equation}
2\xi' \equiv  \phi-\chi_1'+\chi_2', \qquad 2\eta' \equiv  2\rho-\phi-\chi_1'.
\end{equation}
Here we have defined \(R_1'(\beta)^2\) and \(R_2'(\beta)^2\) as
\begin{eqnarray}
R_1'(\beta)^2&\equiv&m_1^2s_1^2c_1^2
-2m_1m_2s_1^2c_1^2\cos2\beta+m_2^2s_1^2c_1^2,
\nonumber\\
R_2'(\beta)^2&\equiv&m_1^2c_1^4+2m_1m_2s_1^2c_1^2\cos2\beta+m_2^2s_1^4,\\
S' \hspace{0.5cm} 
 &\equiv& R_1'(\beta)^2+R_2'(\beta)^2 = m_1^2c_1^2+m_2^2s_1^2. \nonumber 
\end{eqnarray}
\(x_1'\), \(x_2'\) and \(X_3'\) are defined by 
\begin{equation}
x_1'=c_2c_3, \qquad x_2'=s_2s_3c_3, \qquad X_3'=m_3s_2s_3c_3.
\end{equation}
Note that $m_1$ and $m_2$ are included  not in $x_i'$ but in $R_i'$. 
Using the similar arguments before we obtain the constraints for $\beta$  as 
\begin{eqnarray}
\frac{m_1^2 s_1^2 c_1^2 + m_2^2 s_1^2 c_1^2 - g_1'^2}{2m_1m_2s_1^2 c_1^2}
&\le \cos \beta \le&
\frac{m_1^2 s_1^2 c_1^2 + m_2^2 s_1^2 c_1^2 - f_1'^2}{2m_1m_2s_1^2 c_1^2},\\
\frac{f_2'^2 - m_1^2 c_1^4 - m_2^2 s_1^4}{2m_1m_2s_1^2 c_1^2}
&\le \cos \beta \le&
\frac{g_2'^2 - m_1^2 c_1^4 - m_1^2 c_1^4}{2m_1m_2s_1^2 c_1^2}.
\end{eqnarray}
where \(f_1'\) and \(g_1'\) are defined by
\begin{eqnarray}
f_1' &\equiv& \mbox{Max}
\left\{
\begin{array}{c}
 \frac{x_1'(\mm-X_3')-x_2'\sqrt{S'(x_1'^2+x_2'^2)-(\mm-X_3')^2}}
      {x_1'^2+x_2'^2},\\
 \frac{x_1'(X_3'-\mm)-x_2'\sqrt{S'(x_1'^2+x_2'^2)-(\mm-X_3')^2}}
      {x_1'^2+x_2'^2},\\
 \frac{-x_1'(\mm+X_3')+x_2'\sqrt{S'(x_1'^2+x_2'^2)-(\mm+X_3')^2}}
      {x_1'^2+x_2'^2},\\
 0
\end{array}
\right\},\\
g_1' &\equiv& \mbox{Max}
\left\{
\begin{array}{c}
 \frac{x_1'(\mm+X_3')+x_2'\sqrt{S'(x_1'^2+x_2'^2)-(\mm+X_3')^2}}
      {x_1'^2+x_2'^2},\\
 \frac{X_3'+\mm}{x_1'}
\end{array}
\right\}.
\end{eqnarray}
\(f_2'\)(\(g_2'\)) is defined by 
the exchange of \(x_1'\) for \(x_2'\) in \(f_1'\)(\(g_1'\)). 
The constraint which is irrelevant to concrete value of $\beta$ 
is also obtained by changing $x_i$, $R_{1 \pm}$,  $R_{2 \pm}$, $S$  and $X_3$ by 
$x_i'$, $R_{1 \pm}'\equiv|m_2 \pm m_1|s_1c_1$, 
$R_{2 \pm}'\equiv |m_1c_1^2 \pm m_2s_1^2|$, $S'$ and $X_3'$ 
respectively in the previous result Eq.\ (\ref{eq8801}).

\par
In conclusion, we have discussed the constraints of lepton mixing angles from lepton number 
violating processes such as neutrinoless double beta decay, \(\mu^-\)-\(e^+\) 
conversion process by the muonic atom, $\mu^-+N(A,Z+2) \rightarrow e^++N(A,Z)$  
and the lepton number violating K decay, $K^- \rightarrow \pi^+\mu^-\mu^-$, 
which are allowed only if neutrinos are Majorana particles. The rates of 
these processes are proportional to the averaged neutrino mass
defined by $\langle m_{\nu} \rangle_{a b}\equiv |\sum _{j=1}^{3}U_{a j}
U_{b j}m_j|$  ($a,b=e$ and $\mu$) in the absence of right-handed weak coupling.
Here $U_{a j}$ is the left-handed lepton mixing matrix which combines 
the weak eigenstate neutrino ($a=e$ and $\mu$) to the mass 
eigenstate neutrino with mass $m_j$ ($j$=1,2 and 3).   
We obtain the consistency conditions which are satisfied irrelevant to the 
concrete values of $CP$ violation phases (three phases in Majorana 
neutrinos). These conditions constrain the lepton mixing angles, 
neutrino masses $m_i$ and averaged neutrino mass \(\langle m_{\nu} \rangle_{a b}\). 
By using these constraints we have derived the limits on averaged neutrino masses 
\(\langle m_{\nu} \rangle_{\mu e}\) and \(\langle m_{\nu} \rangle_{\mu \mu}\) 
for \(\mu^-\)-\(e^+\) conversion and  K decay, $K^- \rightarrow \pi^+\mu^-\mu^-$, 
respectively.

\ \\
Acknowledgement:

H.N. would like to thank KIAS for the kind hospitality during his visit and 
Chung Wook Kim for various comments on this work.

\clearpage


\clearpage

\vspace{3cm}

\
\\
{\bf Figure Captions}\\
\ \\
{\bf Fig.1:} The Feynman diagrams for 
\ (a) neutrinoless double beta decay, 
\ (b) \(\mu^-\)-\(e^+\) conversion and 
\ (c) K decay, $K^- \rightarrow \pi^+\mu^-\mu^-$.

\ \\
{\bf Fig.2:} The allowed region in the 
$sin^22\theta_1$ versus $\langle m_\nu \rangle_{\mu \mu}^2$ plane obtained 
from \(\mu^-\)-\(e^+\) conversion for the case where $s_3=0$.
The allowed region is given by the shaded areas. 

\ \\
{\bf Fig.3:} The allowed region in the 
$s_1^2$ versus $s_2^2$ plane obtained 
from \(\mu^-\)-\(e^+\) conversion for the case where $s_3=0$.
The allowed region is given by the shaded areas. 

\ \\
{\bf Fig.4:} The allowed region in the 
$s_1^2$ versus $s_2^2$ plane obtained 
from K decay, $K^- \rightarrow \pi^+\mu^-\mu^-$ 
for the case where $s_3=0$.
The allowed region is 
given by the shaded areas in the cases: 
\ (a) $\langle m_\nu \rangle_{\mu \mu} \leq m_1$, 
\ (b) $m_1 \leq \langle m_\nu\rangle_{\mu \mu} \leq m_2 $ and
\ (c) $m_2 \leq \langle m_\nu \rangle_{\mu \mu} \leq m_3 $. 

\ \\
{\bf Fig.5:} The allowed region in the 
$s_2^2$ versus $\langle m_\nu \rangle_{\mu \mu}$ plane obtained 
from K decay, $K^- \rightarrow \pi^+\mu^-\mu^-$ 
for the case where $s_3=0$.
The allowed region is given by the shaded areas. \newline

\clearpage

 \begin{figure}[htbp]
 	\begin{center}
 	\leavevmode
 	\epsfile{file=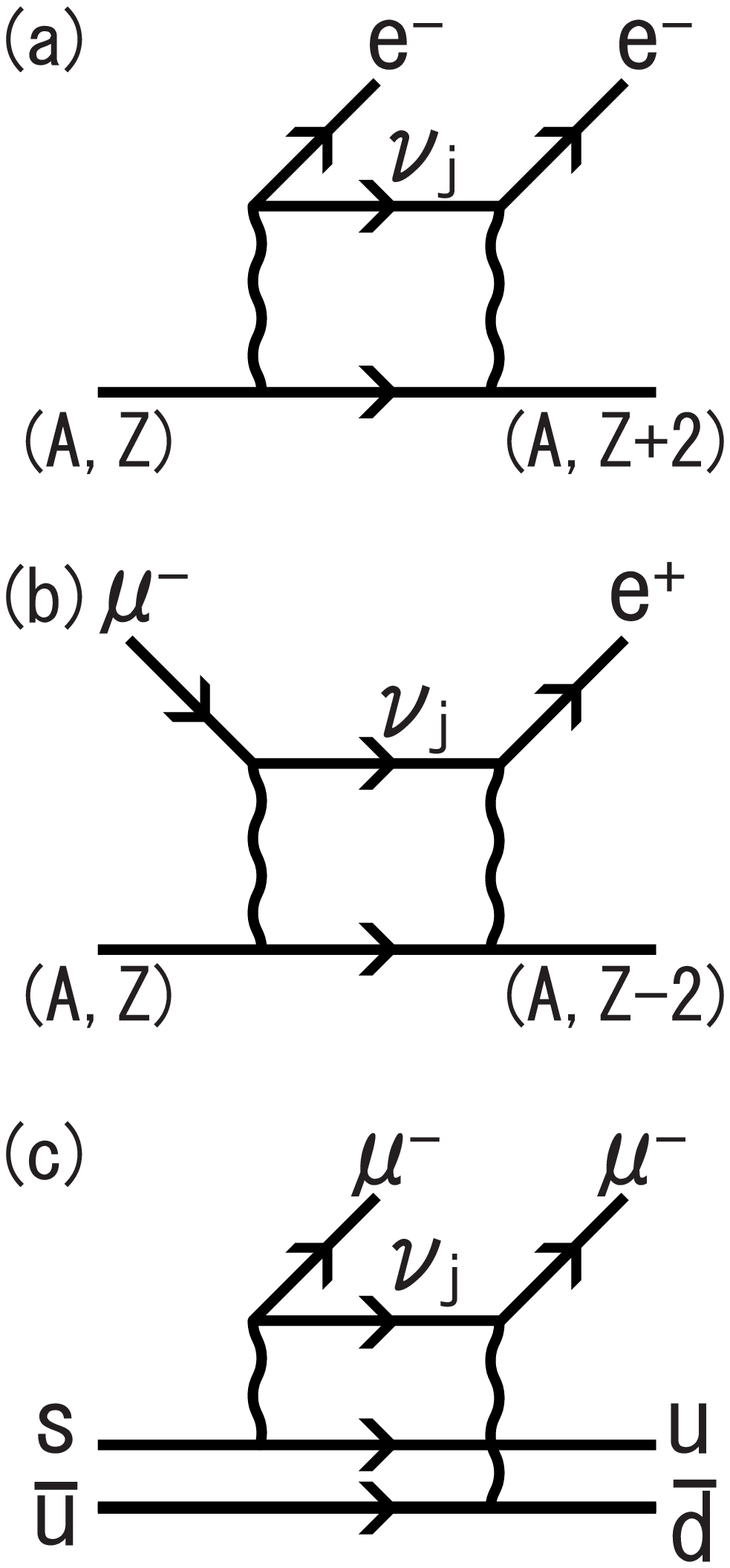,width=8cm}\\
 \ \\
 	{\Huge \bf Fig.1}
 	\end{center}
 \end{figure}

 \begin{figure}[hbtp]
 	\begin{center}
 	\leavevmode
 	\epsfile{file=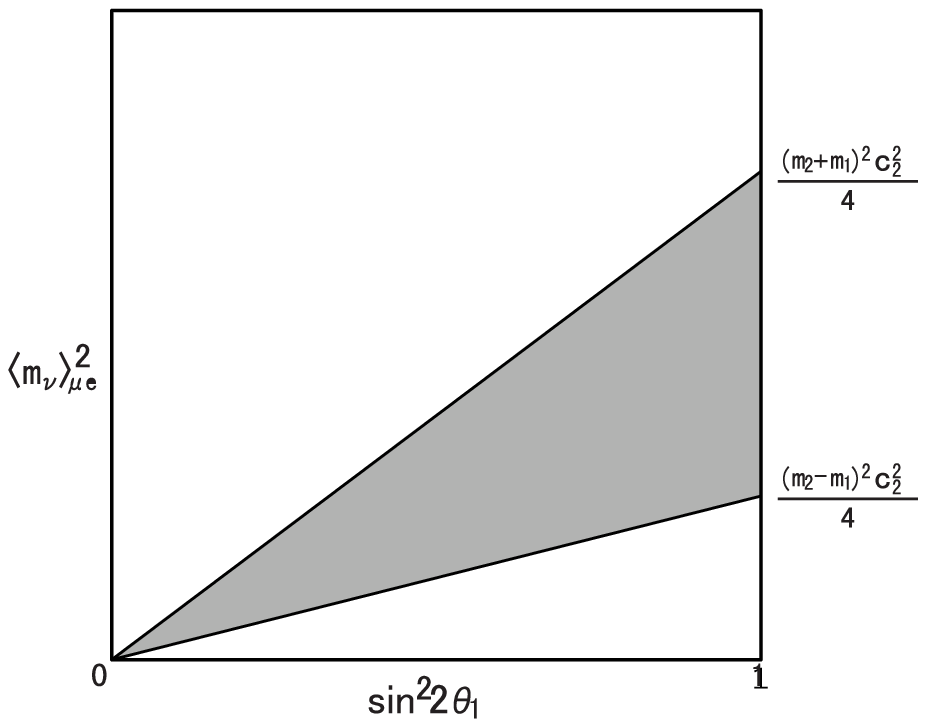}\\
 \ \\
 	{\Huge \bf Fig.2}
 	\end{center}
 \end{figure}

 \begin{figure}[htbp]
 	\begin{center}
 	\leavevmode
 	\epsfile{file=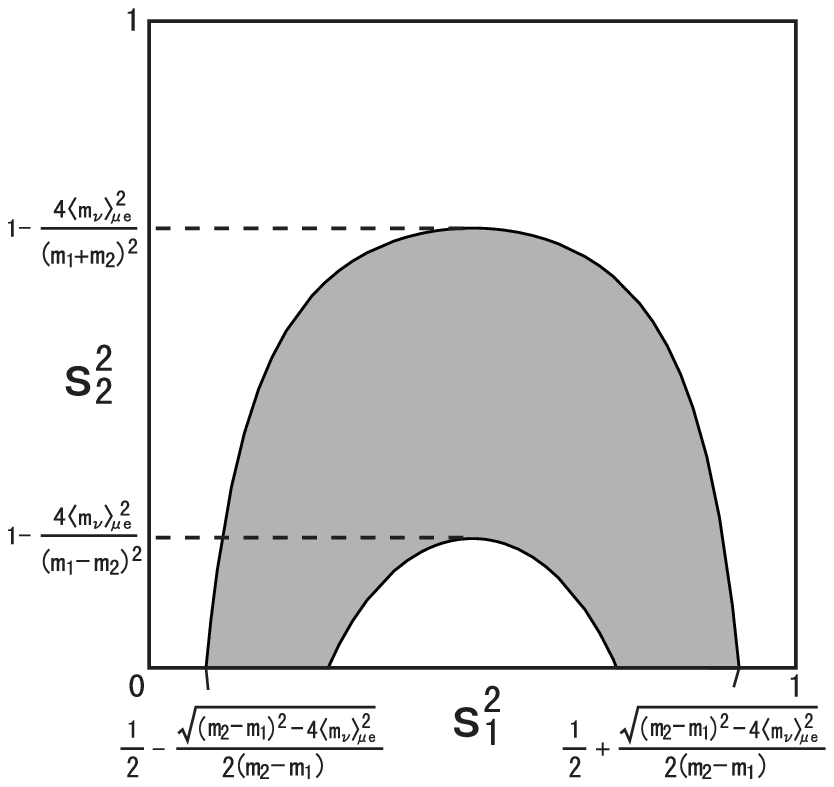}\\
 \ \\
 	{\Huge \bf Fig.3}
 	\end{center}
 \end{figure}

 \begin{figure}[htb]
 	\begin{center}
 	\leavevmode
 	\epsfile{file=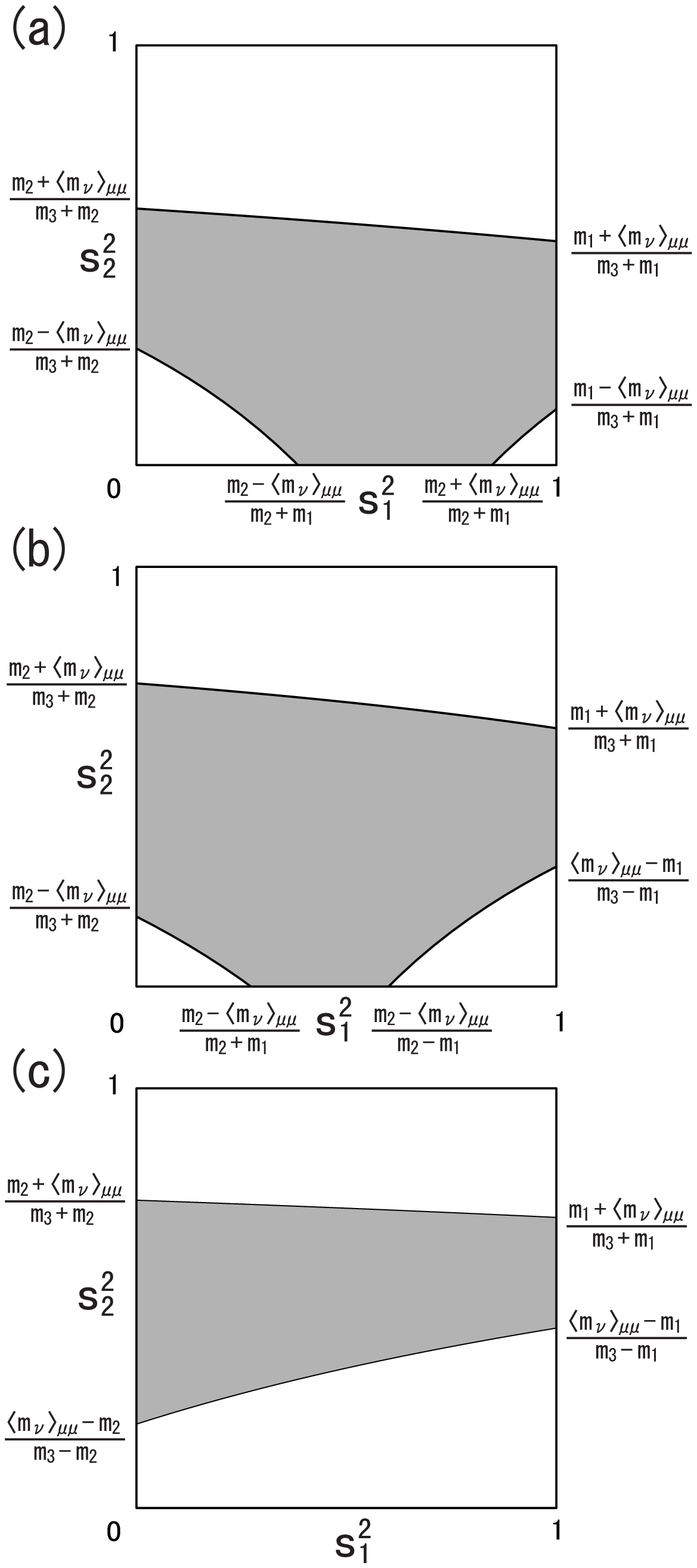,width=8cm}\\
 \ \\
 	{\Huge \bf Fig.4}
 	\end{center}
 \end{figure}

 \begin{figure}[htb]
 	\begin{center}
 	\leavevmode
 	\epsfile{file=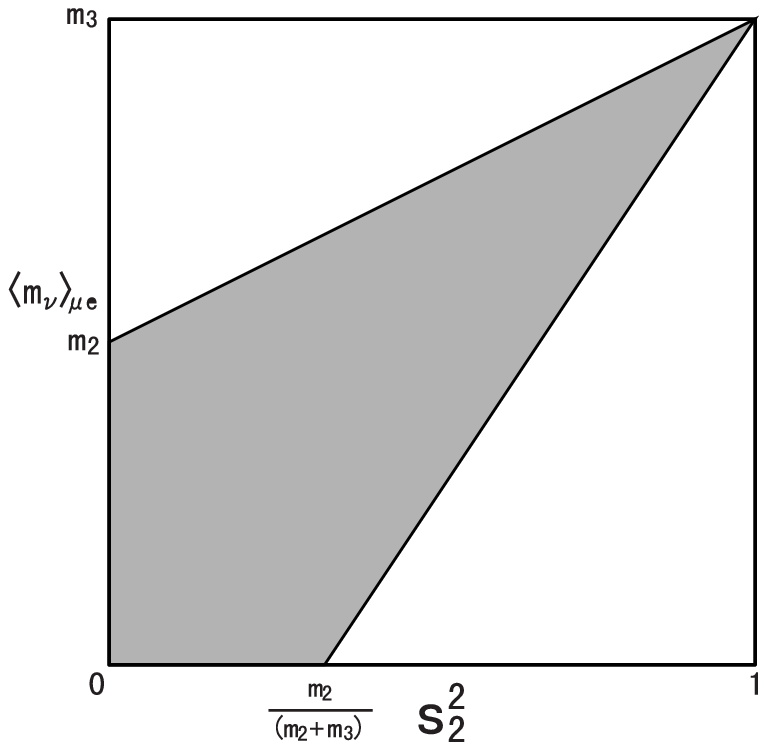}\\
 \ \\
 	{\Huge \bf Fig.5}
 	\end{center}
 \end{figure}
\end{document}